%% file: crypto-identify-icissp2026.tex
\documentclass[a4paper,twoside]{article}

\newif\ifanonymous
\anonymousfalse 

\usepackage[hidelinks]{hyperref}
\usepackage[T1]{fontenc}

\usepackage{subcaption}
\usepackage{calc}
\usepackage{amssymb}
\usepackage{amstext}
\usepackage{amsmath}
\usepackage{amsthm}
\usepackage{multicol}
\usepackage{pslatex}
\usepackage[bottom]{footmisc}
\usepackage{cleveref}
\usepackage{algorithm}
\usepackage{enumitem}
\usepackage{geometry}
\usepackage{graphicx}
\usepackage{fontawesome5}
\usepackage{tikz}
\usetikzlibrary{arrows.meta, shapes.geometric, positioning, shadows, chains}
\usepackage{pgfplots}
\pgfplotsset{compat=1.18}
\usepackage{pgfplotstable}
\usepgfplotslibrary{groupplots, colorbrewer}
\usepackage[normalem]{ulem}
\usepackage{xspace}
\usepackage{booktabs}
\usepackage{todonotes}
\usepackage[american]{babel}
\usepackage{csquotes}
\usepackage{apalike}
\usepackage{url}

\input{misc/misc.tex}
\usepackage{SCITEPRESS}

\begin{document}

\title{Detecting Cryptographically Relevant Software Packages\\with Collaborative LLMs}

\ifanonymous
  \author{Authors and affiliations withheld for Peer Review}
\else
  \author{\authorname{Eduard Hirsch\sup{1}\orcidAuthor{0000-0001-9593-2342}, Kristina Raab\sup{2}\orcidAuthor{0009-0001-7072-0340}, Tobias J. Bauer\sup{2}\orcidAuthor{0009-0006-1073-3971}, and Daniel Loebenberger\sup{2}\orcidAuthor{0000-0002-7969-6260}}
  \affiliation{\sup{1}Ostbayerische Technische Hochschule (OTH), Amberg 92224, Germany}
  \affiliation{\sup{2}Fraunhofer AISEC, Garching 85748, Germany}
  \email{e.hirsch@oth-aw.de, \{kristina.raab, tobias.bauer, daniel.loebenberger\}@aisec.fraunhofer.de}}
\fi


\abstract{
  IT systems are facing an increasing number of security threats, including advanced persistent attacks and future quantum-computing vulnerabilities. The move towards crypto-agility and post-quantum cryptography (PQC) requires a reliable inventory of cryptographic assets across heterogeneous IT environments. Due to the sheer amount of packets, it is infeasible to manually detect cryptographically relevant software. Further, static code analysis pipelines often fail to address the diversity of modern ecosystems. Our research explores the use of large language models (LLMs) as heuristic tools for cryptographic asset discovery. We propose a collaborative framework that employs multiple LLMs to assess software relevance and aggregates their outputs through majority voting. To preserve data privacy, the approach operates on-premises without reliance on external servers. Using over 65,000 Fedora Linux packages, we evaluate the reliability of this method through statistical analysis, inter-model agreement, and manual validation. Preliminary results suggest that~LLM ensembles can serve as an efficient first-pass filter for identifying cryptographic software, resulting in reduced manual workload and assisting PQC transition. The study also compares on-premises and online LLM configurations, highlighting key advantages, limitations, and future directions for automated cryptographic asset discovery.
}

\onecolumn \maketitle \normalsize \setcounter{footnote}{0} \vfill

\section{\uppercase{Introduction}}
\label{sec:introduction}


Current IT infrastructures are constantly under threat from cyber-attacks. Also,  the progression of quantum computing poses a significant risk to existing systems using classical (non-post-quantum) cryptographic schemes \cite{GRI2024QTT}. In addition, no deployed protocol should be assumed secure forever \cite{RFC7696}. Therefore, organizations must be prepared to adapt their cryptographic mechanisms rapidly in response to emerging threats. This capability, known as \textit{crypto-agility} \cite{naether2025cryptographicagility}, is essential for maintaining robust secure services in the face of evolving vulnerabilities, supporting change in cryptographic primitives and protocols.

An initial step toward achieving crypto-agility is the systematic identification of all cryptographic assets (\textit{crypto-assets}) in use. Crypto-assets include key material, cryptographic algorithms, and communication protocols distributed across IT systems. Conducting this process in operational environments is rarely straightforward. Many organizations still lack a complete inventory of their general IT assets, let alone their cryptographic components \cite{EntrustPonemon2024ZTE}. Nevertheless, creating such an inventory is essential. It provides the foundation upon which crypto-agile infrastructures can be developed \cite{CISANISTNSA2023PQCFactSheet}.

Creating a cryptographic asset inventory is especially challenging in large, heterogeneous systems. An organization's software portfolio may span tens of thousands of distinct packages, many of which are linked---directly or through transitive dependencies---to cryptographic libraries. These might be spread over several enterprise systems consisting of thousands of interdependent software packages, container layers, microservices, and embedded components spanning multiple operating systems and hardware generations. Many of these artifacts are upgraded continuously through CI/CD pipelines or package managers. However, others persist on unpatched legacy hosts or in long-lived firmware. Cryptographic functionality may be exposed either explicitly, such as through direct function calls to OpenSSL, or implicitly, buried deep inside transitive dependencies, static libraries, or vendor binaries.

Manually identifying cryptographically relevant software packages is not feasible on a large scale. Traditional static analysis pipelines struggle to handle the breadth and heterogeneity of current software products. The large volume of packages and the version variability make crypto-asset discovery a capacity issue. Thus, (semi-)automatic analysis of these packages is essential.

Against this backdrop, state-of-the-art large language models (LLMs) have emerged as powerful, general-purpose and code-understanding engines. These models are trained using vast amounts of data \cite{openai2023gpt4,meta2024llama3,google2025gemini25pro,anthropic2023claude2card}. This knowledge can be used to identify software packages on large scale that use cryptographic libraries/protocols, or that implement cryptographic algorithms.

This paper investigates if LLMs can assist in the crypto-asset discovery process by accelerating inventory efforts, reducing human error, and laying the groundwork for subsequent cryptographic migration steps \cite{loebenberger2025formalization}. We propose a heuristic-based approach that uses state-of-the-art prompting techniques to query multiple LLMs about the cryptographic relevance of a given package. The study reviews response aggregation via a majority-vote consensus mechanism. Recent literature, \eg\cite{davoudi2025collectiveReasoning,choi2025debateOrVote,li2024moreAgents,kirchhoff2025combiningLlms} shows that, in the absence of ground truth and when human validation is infeasible on a large scale, collaborative knowledge validation through the majority vote of several LLMs provides reliable and efficient approximate reasoning.

Our goal is to create an initial yet actionable inventory of cryptographically relevant software packages to be analyzed further in subsequent PQC migration efforts. This paper addresses the following research questions:

\begin{itemize}[leftmargin=0.85cm, rightmargin=0cm]
  \item[\textbf{RQ1}] How can LLMs be employed to heuristically identify software packages that implement or depend on cryptographic functionality?
  \vspace*{0.25cm}
  \item[\textbf{RQ2}] Is it possible to enhance the quality of responses by utilizing the aggregated outputs of multiple LLM models?
\end{itemize}

The remainder of the paper is structured as follows. In \Cref{sec:related_work}, we review the related literature in this research area. In \Cref{sec:contribution} the open challenges are identified in greater detail that our contribution aims to address. The approach is presented in \Cref{sec:method}, including our methodological setup, the selection of the employed LLMs, and the concrete steps taken to answer our research questions. The statistical evaluation results of our approach are presented in \Cref{sec:results}. The discussion in \Cref{sec:discussion} analyzes our findings and highlights directions for future research. Finally, we conclude the paper with a summary in \Cref{sec:conclusion}.


\section{Related Work}
\label{sec:related_work}

\subsection{Discovery of Cryptographic Functions in Software Packages}

We examined various studies that aim at identifying cryptographic functionality in software packages. Three main approaches recur in the literature: (i)~static code analysis~(STATIC), (ii)~knowledge-driven pattern matching~(KNOW), and  (iii)~LLM-based analysis~(LLM)---often combined in hybrid forms.

STATIC approaches analyze source or binary code without execution (unlike dynamic code analysis), typically using tools such as SonarQube, CodeQL, Ghidra, and IDA Pro. These approaches identify crypto-assets (\eg certificates and keys) or library calls. However, they are often programming language-specific, which limits their range of applications. For example, IBM's CBOM Kit \cite{ibm2025cbomkit} focuses on Java and Python code.

KNOW approaches such as \cite{windriver2020cryptodetector}, \cite{leirimaa_supporting_2024}, and \cite{schmitt_criteria_2024} flag cryptographic functions through expert-curated keyword lists, regular expression rules, or taxonomies. Such approaches often suffer from false positives, dependence on naming conventions, and the need for constant updates. Besides that, the methods are only applicable to specific language setups. For instance, \cite{leirimaa_supporting_2024} is only applicable to JavaScript packages implementing the Node.js crypto module. Consequently, knowledge-driven crypto-asset detection is limited.

We identified two studies that combined STATIC and KNOW elements in their approach. In \cite{rattanavipanon_toolchain_2025} cryptographic functions where detected in Linux binaries by linking a manually compiled list of quantum-vulnerable libraries and functions to static call-graph analysis. Notably, this method analyzes binary executables rather than source code directly, which is advantageous in restricted environments or when third-party assets are involved. However, the method is maintenance-intensive  and error-prone requiring continuous manual updates of quantum-vulnerable libraries and function calls. One final note is that statically linked libraries won't be recognized at all.

Similarly, \cite{meijer_wheres_2021} turn binary code in so-called data flow graphs (DFG). An optimized subgraph isomorphism test then examines the DFG for matching generic signatures (structural patterns) of cryptographic classes that have been pre-defined as taxonomies such as Feistel networks or (N)LFSR---wildcards cover unknown details. Again, this method demands heavy manual effort, thus, not suiting the need of quickly identifying large amounts of cryptographically relevant software packages as aimed for in our approach.

Studies that use STATIC, KNOW, as well as LLM elements complement static and knowledge-driven methods with LLMs. Parsed code syntax trees are checked against a manually curated list of crypto-functions in \cite{moffie_cryptoscope_2025}. To evaluate their approach they create an LLM-based labeled dataset. The labels are generated by prompting the pre-trained models \texttt{ChatGPT-4o} and \texttt{Mistral-Large-2407} to determine whether the provided code contains cryptography. Although promising, the authors note poor quality in the LLM labeling process, thus requiring manual verification. They assume the LLM's performance could be enhanced by more precisely engineered prompts. Furthermore, the approach is limited to Java code. We want to point out that we could not figure out how binary code is analyzed (as depicted in Figure 3 and 4 in \cite{moffie_cryptoscope_2025}) since the paper outlines the analysis of source code only. In addition, no project code is provided further impeding comprehensibility.

The study of \cite{shang2024FoC} fine-tunes an LLM on cryptographic binary summaries generated by ChatGPT, achieving superior binary code interpretation but still depending on a precompiled keyword list of cryptographic function classes.

A review by \cite{fan_rr_2024} examines existing STATIC, KNOW, and LLM approaches for identifying cryptographic functions and concludes that dependence on handcrafted knowledge bases limits scalability and crypto-agility. The authors emphasize compiler variability, obfuscation and optimization strategies, as well as different cryptographic algorithm variants or versions as persistent challenges. They suggest LLMs as promising enablers of adaptive, heuristic detection. Our work directly addresses these limitations by using general-purpose LLMs for heuristic package classification. Additionally, our approach is straightforward and does not require in-depth knowledge of  cryptography or software package analysis, making it accessible to all users.

\subsection{LLM Prompting Techniques}

The prompting techniques proposed in the two subsequent studies, which used LLMs to detect malicious commands in code, significantly influenced our querying method for LLMs. In particular, \cite{zahan_leveraging_2024} apply iterative self-refinement and an \enquote{LLM-as-a-judge} framework to the pre-trained models \texttt{GPT-3} and \texttt{GPT-4}. The authors propose prompting the LLM to respond in JSON format, a method that we found advantageous for producing high-quality responses. Their results show that the LLM outperforms CodeQL-based static analysis.

Static code analysis was combine with LLM reasoning in a two-stage process by \cite{huang_spiderscan_2024}. First, static code analysis is used to depict known malicious commands in graphs. In the second stage, new packages are depicted similarly in graphs, and an LLM is prompted to compare the newly created graph with those from the first stage and classify them as malicious or benign. The authors used advanced prompt design criteria, which we largely adopted. These criteria include explicit task framing, contextual grounding, and JSON-based output requirements, all of which improve interpretability and parsability across different LLMs.

\subsection{Collaborative LLM Reasoning}


When a definitive ground truth is unavailable, aggregating responses from multiple LLMs using a collaborative strategy is effective. Accordingly, \cite{davoudi2025collectiveReasoning} aggregate answers from multiple heterogeneous LLMs via simple majority voting, improving reliability in settings without labeled data or ground truth, provided that prompts are clear and unambiguous. Similarly, \cite{li2024moreAgents} show that majority-vote aggregation across models enhances performance on factual and reasoning tasks. Finally, \cite{choi2025debateOrVote} disentangle majority voting from multi-agent debate across seven natural language benchmarks and find that voting alone accounts for nearly all performance gains, demonstrating that majority voting is a robust, reliable, and accessible collaborative reasoning method.

\section{Our Contribution}
\label{sec:contribution}

In response to the issues and future research directions identified by \cite{fan_rr_2024} and the insights gained from studies on malicious code detection \cite{huang_spiderscan_2024,zahan_leveraging_2024}, we contribute a method that utilizes pre-trained LLMs to identify cryptographically relevant software packages, filling a notable gap in the literature. To avoid violating confidentiality, especially in an enterprise setting, we employ an offline LLM architecture instead of sending queries to online servers. This enables organizations to use our approach for on-premise cryptographic asset discovery. To improve reliability, we further aggregate the individual LLM responses by using a collaborative LLM knowledge strategy based on a majority vote. Our method acts as a preliminary filter to discover cryptographic assets within complex system environments containing numerous software packages and dependencies.

Our method offers several advantages:
\begin{itemize}
  \item It provides an overview of the cryptographic posture of a system's software stack.
  \item The process can be executed with only limited amount of labeled data, which usually can be provided with reasonable effort.
  \item It can be applied even if the local LLMs are less accurate than the latest GPT models, as the majority-vote strategy aggregates the results of multiple models, improving the overall reliability of the classification.
  \item Additionally, our approach is relatively easy to implement, which enhances its practical usability.
\end{itemize}

Our approach provides LLM-derived software package information that can be used to create package dependency graphs and develop a detailed crypto-asset inventory (see \Cref{sec:discussion}). This supports planning for algorithm substitutions as part of PQC migration by enabling more targeted code reviews and focused dependency scanning.

In addition to presenting our practical approach, we examine the plausibility of our method. In \Cref{sec:results:llm_results}, we analyze the LLM responses and the models' level of agreement using statistical methods. Additionally, we manually validate a sample of 390~software packages and use that sample to statistically analyze and infer the performance of the individual models and the majority vote.

For this study, we focus on the Fedora Linux distribution as an example of a complex operating system environment. Fedora is widely used in enterprise settings and offers a rich and stable ecosystem of software packages, making it an ideal candidate for evaluating our approach. However, our method is not limited to Fedora and can be adapted to other Linux distributions or operating systems with similar package management systems.

Based on our analysis of LLM responses, we have developed guidelines for applying the majority-vote strategy to ensure the reliability of LLM-based cryptographic asset discovery. These guidelines, presented in \Cref{sec:results} and \Cref{sec:discussion}, cover best practices for data collection, LLM selection, and response validation in an iterative process.

Finally, we provide all code and data used in this study as open source \cite{OthamiquasyDetectingCryptoPackages2026} to facilitate reproducibility and further research in this area.


\section{Methodology}
\label{sec:method}



\subsection{Approach Overview}

\begin{figure}[ht]
  \centering
  \resizebox{0.9\linewidth}{!}{\input{images/approach.tex}}%

  \caption{Key steps of the cryptographic relevance identification approach.}
  \label{fig:approach}
\end{figure}
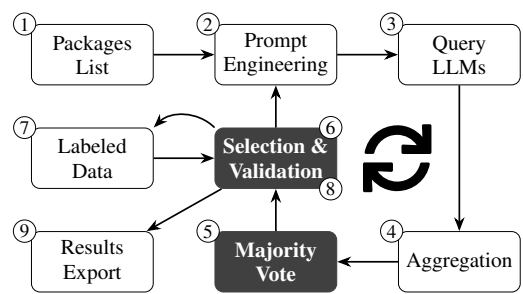

In this section, the iterative approach of this study is presented, illustrated in \Cref{fig:approach}. In \numref{1}, we collect a base package list from a chosen package manager (\Cref{sec:method:data_collection}), which includes the package or module name, version, description, and first-level package dependencies. The package details gathered from the package manager are then used to fill a prompt template, created in \numref{2}. With this prompt, LLMs are queried for determining the cryptographic relevance of each package---step \numref{3}. The prompt is carefully engineered to ensure clarity and effectiveness using state-of-the-art prompt engineering techniques, such as few-shot prompting and instruction prompting. Our prompting is also guided by the findings of \cite{huang_spiderscan_2024,zahan_leveraging_2024}. The resulting prompt is discussed in detail in \Cref{sec:method:model_usage}.

Providing the LLM with the necessary context by including package manager details benefits the approach in two ways:
\begin{enumerate}
  \item Package repositories are typically public and often contain the source code of the packages. We assume this information was used to train commonly available LLM models as mention in the introduction.
  \item Providing additional information, such as descriptions and dependencies, gives LLMs more background knowledge about packages. This allows them to make more informed decisions, even when little to no publicly available information is accessible.
\end{enumerate}

Once a prompt is created, it is applied to $n$ different LLMs, which are selected based on availability and technical diversity (see \Cref{sec:method:model_usage}). The LLMs' responses are then aggregated in \numref{4} followed by \numref{5} a majority-vote strategy. For a set of $n$ models, we classify a package as cryptographically relevant if $\left\lfloor \frac{n}{2} \right\rfloor + 1$ or more models classify it as relevant. This strategy is described in detail in \Cref{sec:method:aggregation}.

Next, a selection process is performed based on the performance matrices of the LLMs' responses. This process identifies the final set of LLMs used for majority voting and employs cross-validation on a subset of packages to validate the results. These steps \numref{6}, \numref{7}, and \numref{8} are explained in \Cref{sec:results:llm_results}. When results are not satisfactory, the process can be iterated by refining the prompt. This iterative approach allows for continuous improvement of the classification accuracy. The final results in \numref{9} are then exported for further processing, \eg to build a first version of a cryptographic bill of materials (CBOM).

Additionally, to the workflow itself, patterns and correlations are examined (\Cref{sec:results:distribution}) helping to assess the plausibility of the results and identify any potential issues. The overall results are presented in \Cref{sec:results} where the complete assessment is shown in detail. It illustrates the accuracy of the chosen models, classifying the range of quality to expect when applying the approach in practice.

\subsection{Data Collection}
\label{sec:method:data_collection}


For each software package basic specifications are publicly available, including the name, description, and first-level dependencies. This information is crucial for prompting an LLM to make useful decisions about a package's cryptographic relevance. Common package managers, such as \textit{dnf, apt, pip,} and \textit{npm} can retrieve this information from a system environment. There are also tools that can detect the operating system or package repository and extract the necessary information, such as the open source tool Syft\footnote{Syft: \url{https://perma.cc/PXU5-SR99}}.

For this study, the \textit{dnf}\footnote{DNF: \url{https://perma.cc/W45V-H8QQ}} package manager from the Fedora Linux distribution has been used to compile a list of packages. Most of the extracted packages were fetched from the official Fedora and RPM Fusion repositories, but we also included packages from other sources, such as Docker, VirtualBox, and Visual Studio Code. We where trying to cover as many packages as possible to get a good picture of the cryptographic relevance for one particular ecosystem.


After exporting a package list, we removed version- and architecture-specific information from package names (\eg \texttt{openssl\-3.2.1\-1.fc40.x86\_64} was mapped to \texttt{openssl}). Further, the package description is picked from the latest available version. We assume that package descriptions remain largely consistent across versions with respect to cryptographically relevant information. Nevertheless, when fetching packages from a running system, it would be preferable to grab descriptions from the installed version.

However, considering version details is crucial in a later crypto-asset discovery stage, since this information is necessary for determining which specific algorithms are supported or implemented, as well as for identifying any vulnerabilities.

In summary, the final deduplicated list contains 65,295 packages that will be used to query the LLMs in a subsequent step.

\subsection{Model Usage}
\label{sec:method:model_usage}

We deployed five different local LLMs, which are freely available and can be run on local machines to preserve privacy. For good performance, however, a powerful GPU is recommended; otherwise, the inference time can be quite long. To put this into relation---without going into the detailed specifications of the machines---a single query on a developer notebook took, at the time of writing, about 4 minutes and on a powerful DGX machine about 2 to 20 seconds. For technical details on the hardware and software setup, see \Cref{sec:method:experimental_setup}.

The five LLMs have been selected for their availability, fast access, and technical diversity. The models were hosted using two different infrastructures: \gptFall\footnote{\gptFall: \url{https://perma.cc/R3NY-S7AY}} and \ollama\footnote{\ollama: \url {https://perma.cc/Y9JB-CTNQ}}. \gptFall is a framework that allows querying various LLMs using a Python API without running a dedicated server. \ollama, in contrast, provides a server which allows accessing models over a REST API. This setup enables us to leverage the strengths of both infrastructures and models.

For \gptFall following models are included:
\begin{itemize}
    \item \texttt{gpt4all-13b-snoozy-q4\_0}
    \item \texttt{Meta-Llama-3-8B-Instruct.Q4\_0}
    \item \texttt{Nous-Hermes-2-Mistral-7B-DPO.Q4\_0}
    \item \texttt{Phi-3-mini-4k-instruct.Q4\_0}
\end{itemize}

On the \ollama server, we used \texttt{DeepSeek R1}, which facilitates an iterative thinking processes to improve response quality.

The prompt texts have been designed to provide the LLM with the necessary package information. For each package, specific placeholders are filled with the actual values from the package list, including the name, description, and dependencies. Then, the LLM is queried about the cryptographic relevance of the package in question. To ensure consistent LLM responses, the prompt requires a specific output format (JSON).

Initially, the same prompt was used for all models. However, during the evaluation process, we found that modifying the prompt improved the results for some models. Therefore, we adapted the prompt for certain models.

With this approach, LLMs are given clear instructions on how to respond, employing techniques such as one-shot and instruction prompting. We also experimented with different approaches, such as providing a specific package as a one-shot example or creating a longer list of examples. However, we found that one-shot prompting with an output template using descriptive names as field specifications produces the best results. By \enquote{best results}, we mean that the LLMs are able to output a parsable JSON response, which is necessary for the subsequent aggregation step. In summary, keeping the prompt very short and precise is key to success.

\subsection{Aggregation}
\label{sec:method:aggregation}


All queried LLMs where instructed produce consistent JSON responses, which are then parsed and stored in a CSV file. The retrieved JSON object contains three attributes: \enquote{package}, \enquote{cryptographic\_relevance}, and \enquote{justification}. It is important to note that the responses are not always in the correct JSON format. This can occur for various reasons, such as missing quotation marks, or brackets, or missing answer fields altogether. These issues persisted when we attempted to optimize the prompt further. Therefore, we implemented a JSON parser that can handle common errors and anticipate these issues. For instance, the parser corrects missing or incorrect quotation marks and brackets, as well as slightly different output formats across different models. However, sometimes the response is malformed and cannot be parsed at all. In these cases, we simply discard the response and add a row to the CSV file containing the name and empty fields. In \Cref{sec:results}, we present the number of non-parsable LLM answers in relation to the parsable ones.

Not just precise prompts but also invocation parameters are important for generating correct and reproducible responses. Parameters, such as \texttt{temperature}, \texttt{top-p}, and \texttt{max\_tokens} can significantly influence the output of the LLMs. Where possible, we set the \texttt{temperature} to 0, aiming for near-deterministic outputs. However, to boost model independence it might be beneficial to increase temperature.

Other parameters were left at the model defaults. Specifically, we kept \texttt{top-p} at the model defaults (0.4 for \gptFall and 0.95 for \ollama) to preserve the intended balance between determinism and variability \cite{arora2024optimizing}. We hypothesize that some variability could reduce correlations and prevent different models from producing nearly identical responses. This is particularly important for the majority-vote strategy. Otherwise, it would decrease the effectiveness of the aggregation process (see \Cref{sec:results}).


There were a few minor issues with the CSV files that needed to be addressed, such as non-ASCII characters and malformed entries. These issues were manually corrected or removed, resulting in clean CSV files. Finally, we verified the files by a CSV linter. Each final CSV file per model contains the following information for each package: the package name, the cryptographic relevance classification (String "True" or "False"), and the justification provided by the LLM.

Next, all LLMs' CSV files were merged into one \textit{pandas} data frame for evaluation. The consolidated data frame contains for every package (rows) the classification results from each LLM (columns). Besides the individual classifications, an aggregated decision is provided in an extra column. This aggregation is based on the majority-vote strategy. As previously mentioned, if at least $\left\lfloor \frac{n}{2} \right\rfloor + 1$ models agree on a classification, the package is classified as such. For instance, if three out of the five LLMs classify a given package as cryptographically relevant, the majority vote is considered \texttt{true} for that package.

\subsection{Selection and Validation}
\label{sec:method:selection_and_validation}

We selected a representative sample of 390 packages to serve as the ground truth against which to validate the results of our approach. The sample was generated by randomly selecting packages from the six strata, grouped according to the number of \texttt{true} counts (ranging from~0~to~5). This stratification ensured a diverse range of packages, while also accounting for the varying levels of agreement among the models. The package details---such as descriptions and dependencies---were examined manually and labelled according to their actual cryptographic relevance.

We used the labelled sample to validate our approach iteratively with performance metrics (see \Cref{sec:results:verification:assessment} and \Cref{sec:results:reiteration}). In addition, a cross-validation was performed to assess the reliance on the selected models and the overall performance of the majority-vote strategy. Based on the cross-validation, a set of models was selected (see \Cref{sec:results:final_optimization}).

\subsection{Experimental Setup}
\label{sec:method:experimental_setup}


The described process of \Cref{fig:approach} requires managing three important technical aspects:
\begin{itemize}
  \item \textit{data collection} of the package list,
  \item \textit{model execution} to query the LLMs, and
  \item \textit{result aggregation and evaluation} of the LLM responses.
\end{itemize}

\textit{Data collection} can be performed on any machine with access to the selected package manager and requires no special hardware. Software requirements depend on the chosen collection process; in our case, we used the \textit{dnf} package manager on a Fedora Linux distribution.

\textit{Model execution} requires a powerful machine with a GPU support to efficiently run LLMs, especially when covering numerous packages. However, if the number of packages is very small, a standard machine with a lightweight GPU may suffice. Nevertheless, when using non-specific hardware, interference times of around four to five minutes per package can be expected. In this study, a DGX machine with seven NVIDIA A100 GPUs was used to run five LLMs, each with around 65,000 to 70,000 queries.

The models were executed in rootless Docker containers using the Python integration of \gptFall, while the DeepSeek R1 model was served via an \ollama server. Each \gptFall container was assigned one A100 GPU, allowing parallel execution, and the \ollama server likewise ran on a single A100 GPU. Queries to \gptFall were issued directly through its Python API, which is significantly faster than the REST-based \ollama interface. As a result, all \gptFall runs completed in approximately one day, whereas the \ollama server required about three days.

\textit{Result aggregation and evaluation} was performed in a Jupyter notebook, where we processed the aggregated CSV outputs in a data frame. In addition to descriptive statistics (see \Cref{sec:results:llm_results}), we manually verified our results using a representative set of packages and optimized prompt specifics. We also compared the results of the offline LLMs with those of online models (see \Cref{sec:results:verification}).

\section{Results and Verification}
\label{sec:results}

\subsection{Overview of the Package Classification Across LLMs}
\label{sec:results:llm_results}

\subsubsection{Analysis of LLM Response Quality}

First, we evaluated each LLM by the parsability of its responses. If an answer for a given package cannot be parsed as a valid JSON object---even when common errors, such as missing quotes or brackets, are anticipated---it is marked as invalid. Note that content-related issues, such as missing fields, misspelled attributes, and completely malformed answers, are also counted as erroneous and therefore invalid. \Cref{tab:packages_unanswered} shows the response quality as the number of valid answered responses compared to the number of erroneous responses for each model. Further, model size does not directly correlate with the number of errors. For example, the smallest model \texttt{phi} (2.1~GB) produced the fewest errors (72), while the largest model \texttt{gpt4all} (6.9~GB) produced the most (1,137).

\begin{table}[ht]
  \centering
  \caption{Valid and invalid packages responses per model}
  \label{tab:packages_unanswered}
  \small
  \begin{tabular}{lrrrrr}
    \toprule
    \textbf{model}  & \textbf{size} & \textbf{valid} & \textbf{invalid} & \textbf{error-rate} \\
    \midrule
    phi             & 2.1 GB        & 65,222         & 72               & 0.11\%\\
    deepseek        & 5.2 GB        & 65,199         & 95               & 0.15\%\\
    llama           & 4.4 GB        & 65,094         & 200              & 0.31\%\\
    mistral         & 3.9 GB        & 64,974         & 320              & 0.49\%\\
    gpt4all         & 6.9 GB        & 64,157         & 1,137            & 1.74\%\\
    \midrule
    agg             & ---   & 63,529         & 1,765   & 2.70\%\\
    \bottomrule
  \end{tabular}
\end{table}

It is important to mention that the total number of packages with errors is less than the sum of the individual models because some packages were not answered by multiple models. The total number of unanswered packages is 1,765 (about 2.70\% of all packages). We discarded packages with at least one error in the LLM responses from the dataset for further analysis. This ensured that only packages for which all five models returned a valid answer were included. Consequently, the cleaned dataset contains a total of 63,529 packages.

Examining the individual models reveals room for improvement in the answer rates. For example, the \texttt{gpt4all} model had the most unanswered packages (1,137)---error logs showed that the JSON output could not be parsed in many cases. However, many of these issues stemmed from consistent formatting errors that could have been corrected. For the scope of this paper, we decided not to apply further error correction and to maintain the original results.

Several additional strategies could be applied. Beyond improving the answer parser, one could re-prompt the LLMs for packages that produced errors. This may involve re-querying the models with the same prompt but a slightly higher temperature to introduce variability and potentially avoid repeated failures. Alternatively, the prompt could be revised to improve clarity and reduce ambiguity. These strategies where applied during re-iteration of the approach, as described in \Cref{sec:results:reiteration}.

\subsubsection{Analyzing Data Distribution}
\label{sec:results:distribution}

\begin{figure}[ht]
    \centering

    \hspace*{-2mm}
  \resizebox{\linewidth}{!}{\input{images/chart-empirical-vs-binom-vs-betabinom.tex}}%

    \caption{Package distribution by number of \texttt{true} votes across LLMs compared to empirical, binomial, and beta-binomial distribution.}
    \label{fig:true_votes}
\end{figure}
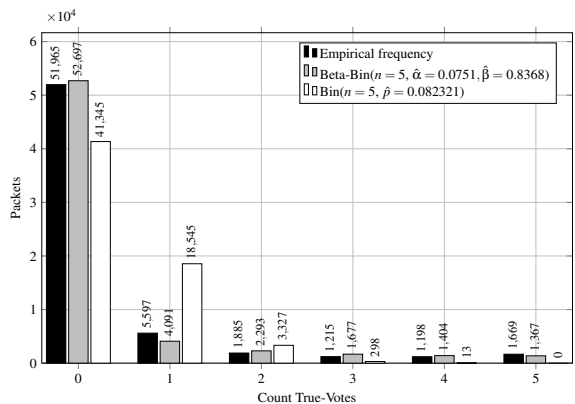

The valid responses of the LLMs were aggregated based on the number of LLMs that voted \texttt{true} (cryptographically relevant). This is illustrated by the black bar in \Cref{fig:true_votes}. Specifically, the $x$-axis categorizes the agreement level, ranging from 0 to 5. This indicates how many LLMs voted \texttt{true} for a particular package. The $y$-axis shows the number of packages that fall into the respective agreement level. For instance, 1,669 packages were deemed cryptographically relevant by all five models.

To better understand the results, the empirical distribution of the \texttt{true} votes was first matched to a binomial distribution. As shown by the white bar in \Cref{fig:true_votes}, there are significant deviations in the empirical distribution. The chi-squared goodness-of-fit test proves this (joining groups 4 and 5 achieves a frequency greater than 5), where $\chi^2 \approx 612,668$ and the number of degrees of freedom $df = \text{\#number-of-groups} - 1 - \text{\#estimated-parameters} = 5 - 1 - 1 = 3$. Looking up the critical value for $\chi^2$ with $df=3$ in a $\chi^2$ distribution table\footnote{~$\chi^2$ distribution table: \url{https://perma.cc/P2UC-U63Q}}, we find that the calculated value of $\chi^2 \approx 612,668$ is much larger than the critical value of $16.266$ in the distribution table at significance level $\alpha = 0.001$. As a result, we can reject the null hypothesis that the empirical data follows a binomial distribution at a 0.1\% significance level. In other words, the probability of observing such a large deviation by random chance is less than 0.1\%.

Second, the empirical data were fitted with a beta-binomial distribution. A beta-binomial distribution is a discrete probability distribution that arises when the probability of success in a series of Bernoulli trials follows a beta distribution instead of being fixed. This allows for modeling overdispersion in the data, which is evident in our data because the assumption of independent and identically distributed trials does not hold. As shown by the gray bar in \Cref{fig:true_votes}, the beta-binomial distribution appears to fit the empirical data better than the basic binomial distribution.

However, the chi-squared goodness-of-fit test gives us now $\chi^2 \approx 767.63$ with $df = 5 - 1 - 2 = 2$ (two estimated parameters). Looking up the critical value for $\chi^2$ with $df=2$ in the same $\chi^2$ distribution table as before, we see that the critical value $13.82$ at significance $\alpha = 0.001$ is still far off.

The two parameters of the beta-binomial distribution are insufficient to fully capture the empirical voting behavior; the data remains considerably overdispersed. This suggests that the probability $p$ of a model voting \texttt{true} is not constant---neither across LLMs nor across packages.

The persistent overdispersion implies that the independence assumptions of the binomial framework does not hold, reflecting latent variability across LLMs, packages, or their interaction. The next section investigates these dependencies in detail.

\subsubsection{Independence Assumption of LLMs}
\label{sec:results:llm_independence}


For the majority-vote strategy to be effective, the individual models must be as independent as possible. However, in practice, this assumption may not hold true in practice. There are several reasons why the models may exhibit correlations \cite{kim2025correlated} in their responses:

\begin{itemize}
  \item Shared training data sources
  \item Similar architectures and objectives
  \item Tuning based on human feedback
  \item Shared benchmarks and evaluation loops
  \item Linguistic and conceptual regularities
  \item Model averaging effects in large-scale training
  \item Systematic biases
\end{itemize}

To understand the dependencies between the models, we introduce the concept of the \texttt{design effect} and the \texttt{effective sample size}. The \texttt{design effect} \cite{Kish1965,DonnerKlar2000} expresses the increase in sampling variance that occurs when observations are more correlated than independent. For a cluster of size $n$ with an average inter-class correlation $\rho$ between the observations, the design effect $\operatorname{D}_{\text{eff}} $ is calculated as follows:
\begin{equation}
\operatorname{D}_{\text{eff}} = 1 + (n - 1) \cdot \rho
\label{eq:design_effect}
\end{equation}

The \texttt{effective sample size} (ESS) is then defined as the size of an equivalent independent sample that would provide the same level of precision as the correlated sample. It can be calculated using the design effect as follows:
\begin{equation}
n_{\text{eff}} = \frac{n}{\operatorname{D}_{\text{eff}} }
\label{eq:effective_sample_size}
\end{equation}

As the inter-class correlation $\rho$ is not yet defined, we can use an approximation. According to \cite{DonnerKlar2000}, the design effect can be estimated by the relation of the observed $\operatorname{Var}_{\text{obs}}$ to the independent variance $\operatorname{Var}_{\text{ind}}$:
\begin{equation}
\operatorname{D}_{\text{eff}}  = \frac{\operatorname{Var}_{\text{obs}}}{\operatorname{Var}_{\text{ind}}}
\end{equation}

Given that we have five models ($n=5$), $\operatorname{Var}_{\text{ind}}=0.3777$ (based on Bin($n=5$, $\hat p=0.082321$)) and $\operatorname{Var}_{\text{obs}}=1.17$ (empirical data), we can estimate the design effect and effective sample size as follows:
\begin{align}
\operatorname{D}_{\text{eff}}  &= \frac{1.17}{0.3777} \approx 3.096 \\
n_{\text{eff}} &= \frac{5}{3.096} \approx 1.615
\end{align}

This indicates that, despite evaluating each package with five models, the combined responses provide information equivalent to roughly \textbf{1.6 independent evaluations}. In other words, the model outputs are moderately to strongly correlated ($\rho \approx 0.52$), reflecting a substantial degree of shared bias or consensus among the models.


\subsection{Results Verification}
\label{sec:results:verification}

\subsubsection{Manual Assessment}
\label{sec:results:verification:assessment}

As shown in \Cref{sec:results:distribution}, the true votes across the LLMs are not uniformly distributed. Therefore, some packages appear to be more difficult to classify than others. To ensure that our manual assessment covers all levels of classification difficulty, we employed a stratified sampling approach. For each agreement level from 0 to 5 (see \Cref{fig:true_votes}), a random sample of 65 packages was selected, resulting in a total of 390 packages. The manual assessment was performed by reviewing the package descriptions and dependencies. When necessary, we also reviewed the source code and documentation. Based on this information, we determined the cryptographic relevance of each package. The sample size of 390 packages is based on a 95\% confidence level and a 5\% margin of error for the 63,529 packages.

\begin{table}[ht]
\centering
\caption{Performance metrics for selected LLMs on Fedora security relevance classification.}
\label{tab:llm_performance_local}
\small
\begin{tabular}{lcccccc}
\toprule
\textbf{Model} & \textbf{Acc} & \textbf{Spec} & \textbf{Prec} & \textbf{Recall} & \textbf{F1} \\
\midrule
\texttt{majority}  & 0.73 & 0.77 & 0.74 & 0.70 & 0.72 \\
\texttt{llama}     & \textbf{0.76} & 0.69 & 0.72 & 0.83 & \textbf{0.77} \\
\texttt{phi}       & 0.72 & 0.54 & 0.65 & \textbf{0.92} & 0.76 \\
\texttt{mistral}   & 0.75 & 0.87 & 0.82 & 0.62 & 0.70 \\
\texttt{deepseek}  & 0.67 & 0.71 & 0.67 & 0.64 & 0.66 \\
\texttt{gpt4all}   & 0.67 & \textbf{0.94} & \textbf{0.86} & 0.38 & 0.53 \\
\bottomrule
\end{tabular}
\end{table}

\Cref{tab:llm_performance_local} presents the performance metrics for the chosen local models and the majority-vote (\texttt{majority}) strategy. These metrics include accuracy (Acc), specificity (Spec), precision (Prec), recall, and F1-score. In this first iteration, both the models and the majority-vote strategy perform rather poorly. The best-performing model is \texttt{llama}, which outperforms the majority vote with an F1-score of 77\%. These results indicate significant room for improvement in the models' classification performance.

In summary, we found that the majority vote averages out performance. While this mitigates the effect of weak models, it also suppresses the strengths of better ones.

\subsubsection{External Model Verification}

To compare the results of the local models with those from state-of-the-art, cloud-based models, we queried four LLMs from well-known providers using the same sample set of 390 packages. These included from OpenAI \texttt{gpt-5}, Google Gemini \texttt{2.5 flash} and \texttt{2.5 pro} versions, and last from Mistral AI the model \texttt{codestral-2508}.

After the initial evaluation as described in the previous section, the set of 390 packages was regarded as ground truth for evaluating the performance of the online models. We used the same prompt and majority-vote strategy, and calculated the same performance metrics. The results are summarized in \Cref{tab:llm_performance_online}.

\begin{table}[ht]
\centering
\caption{Performance metrics for chosen online models.}
\label{tab:llm_performance_online}
\small
\begin{tabular}{lcccccc}
\toprule
\textbf{Model} & \textbf{Acc} & \textbf{Spec} & \textbf{Prec} & \textbf{Recall} & \textbf{F1} \\
\midrule
\texttt{majority}   & \textbf{0.86} & 0.84 & 0.84 & 0.87 & \textbf{0.86} \\
\texttt{openai}     & 0.84 & 0.77 & 0.79 & 0.91 & 0.85 \\
\texttt{geminiflash}& 0.84 & 0.73 & 0.77 & 0.94 & 0.85 \\
\texttt{geminipro}  & 0.81 & 0.65 & 0.72 & \textbf{0.97} & 0.83 \\
\texttt{codestal}   & 0.82 & \textbf{0.90} & \textbf{0.88} & 0.74 & 0.80 \\
\bottomrule
\end{tabular}
\end{table}

The table shows that the online models perform better than the local ones, but only by 3\% to 9\% in terms of F1-score (the best local model is \texttt{llama}, with an F1-score of 77\%). Furthermore, \texttt{majority} seems to work better with the online models than with the local ones. Since our approach favors local LLMs for privacy reasons, we used the online models only for external model comparison.

\subsubsection{Re-iteration and Prompt Optimization}
\label{sec:results:reiteration}

The ground truth generated by the manual validation in \Cref{sec:results:verification:assessment} was used to run the iterative process in \Cref{fig:approach}. During that, we first refined the parsing logic to tolerate common formatting errors.  Second, a re-querying of the models using increased temperature has been implemented if the response could not be parsed correctly. Third, we optimized the prompt for each local model based on the first evaluation round's findings. These optimizations included adjusting the wording, providing clearer instructions, and tailoring the prompt to better suit the respective model's strengths. Larger models, such as \texttt{deepseek}, benefited from longer, more detailed prompts, whereas smaller models, such as \texttt{phi}, worked better with shorter, more concise ones.

\begin{table}[h]
\centering
\caption{Performance metrics for selected local models after prompt optimization.}
\label{tab:llm_performance_local_optimized}
\small
\begin{tabular}{lcccccc}
\toprule
\textbf{Model} & \textbf{Acc} & \textbf{Spec} & \textbf{Prec} & \textbf{Recall} & \textbf{F1} \\
\midrule
\texttt{majority} & \textbf{0.85} & 0.75 & 0.78 & \textbf{0.95} & \textbf{0.86} \\
\texttt{deepseek} & \textbf{0.85} & 0.88 & \textbf{0.87} & 0.81 & 0.84 \\
\texttt{llama}    & 0.79 & 0.80 & 0.79 & 0.78 & 0.78 \\
\texttt{phi}      & 0.72 & 0.55 & 0.65 & 0.90 & 0.76 \\
\texttt{mistral}  & 0.78 & 0.91 & \textbf{0.87} & 0.64 & 0.74 \\
\texttt{gpt4all}  & 0.67 & \textbf{0.94} & 0.86 & 0.39 & 0.53 \\
\bottomrule
\end{tabular}
\end{table}

As shown in table \Cref{tab:llm_performance_local_optimized}, the prompt optimization significantly improved the performance of the local models. The majority-vote strategy outperforms all individual models, achieving an F1-score of 86\%. The best individual model is now \texttt{deepseek} with an F1-score of 84\%, followed by \texttt{llama} with 78\%. These results demonstrate the crucial role of prompt engineering in enhancing LLMs' performance for specific tasks, such as cryptographic relevance identification. The offline-generated majority vote is now competing with online models (see \Cref{tab:llm_performance_online}).

\subsubsection{Final Optimizations}
\label{sec:results:final_optimization}

Finally, based on the findings from the previous sections, we aim to further optimize the majority vote. Substituting the expression $\operatorname{D}_{\text{eff}}$ into \Cref{eq:effective_sample_size}, and using the estimated intra-class correlation $\rho \approx 0.52 \approx \frac{1}{2}$, yields the following expression for the effective sample size:
\begin{align}
n_{\text{eff}} = \frac{n}{1 + \frac{(n - 1)}{2}} &= \frac{2n}{n + 1} \text{ with } \\
 \lim_{n \to \infty} \frac{2n}{n + 1} &= 2
\end{align}

As more models are added to the ensemble, the effective sample size increases only slightly. With three models, it is about 1.5; with five, it rises only to about 1.67. The curve then levels off and approaches a maximum of 2. Thus, using three to five models captures most of the possible gain (given $\rho \approx 0.52$). Adding more models offers little benefit unless they are much more independent, which is difficult to achieve in practice (see \Cref{sec:results:llm_independence}).

Returning to the approach in \Cref{fig:approach}, the validation step first identifies the three best models for the final majority-vote strategy. For the purposes of this study, this is accomplished by evaluating all possible combinations of three models from the set of five local models. The performance metrics for each combination are calculated based on the ground truth established in \Cref{sec:results:verification:assessment}. This can be done in various ways, \eg by optimizing for F1-score, recall, or by a custom weighted metric.

Next, we reduce the selection to three out of five models, which only slightly decreases $n_\text{eff}$ but allows the model ensemble to eliminate non-performing models while ensuring there are enough models to form a stable majority vote. To focus on crypto\-graphically relevant packages, we are aiming for a weighted approach favoring recall, denoted by $r$ over specificity, denoted by $s$. Thus, we optimized for $\operatorname{Score}$ defined as follows:
\begin{equation}
  \operatorname{Score} = w_\text{r} \cdot r + w_\text{s} \cdot s
\end{equation}

with weights $w_\text{r} = 0.7$ and $w_\text{s} = 0.3$. This choice reflects the importance of correctly identifying cryptographically relevant packages (high recall) while still maintaining a reasonable level of accuracy in identifying non-relevant packages (specificity). The models found during that optimization are \texttt{deepseek}, \texttt{phi}, and \texttt{mistral}.

To estimate the internal generalization performance of the majority-vote ensemble, a stratified five-fold cross-validation (CV) was conducted on the labeled validation set. In each of the five iterations, the optimal three-model ensemble was selected on the training subset and subsequently evaluated on the held-out fold. \Cref{tab:cv_results} summarizes the averaged metrics across all folds. The optimal set was calculated based on the weighted $\operatorname{Score}$ as described above. Using the cross-validation the ensemble achieved an average F1-score of 0.82, with consistently high recall (0.85) and balanced precision and specificity. This indicates a stable performance across the splits.

\begin{table}[h!]
\centering
\caption{Five-fold cross-validation results for the majority-vote ensemble. Mean $\mu$ and standard deviation $\sigma$ are reported across folds.}
\label{tab:cv_results}
\begin{tabular}{lccccc}
\toprule
\small
\textbf{Metric}    & \textbf{Acc} & \textbf{Spec} & \textbf{Prec} & \textbf{Recall} & \textbf{F1} \\
\midrule
\textbf{$\mu$}     & 0.82         & 0.79          & 0.79          & 0.86            & 0.82 \\
\textbf{$\sigma$}  & 0.056        & 0.046         & 0.050         & 0.075           & 0.059 \\
\bottomrule
\end{tabular}
\end{table}

The small variation across folds indicates that the ensemble is consistent and is not overly influenced by specific subsets of the data. Despite the limited set of labeled data (390 samples), cross-validation is a practical method for assessing the reliability of these results. Overall, the outcomes suggest solid internal generalization of the ensemble approach. However, the results still reflect internal validation rather than evaluation on an independent test set.

\section{Discussion}
\label{sec:discussion}


As the results in \Cref{sec:results} demonstrate, local LLMs can effectively identify cryptographically relevant software packages when combined in a majority-vote strategy. Several optimizations were essential to achieving these results. First, prompt engineering played a significant role. Tailoring prompts to the strengths of each model and keeping them concise improved response quality. Second, refining the parsing logic to handle common formatting errors allowed for the use of more valid responses. Third, selecting models based on their complementary strengths---such as high recall or specificity---enhanced the overall performance of the majority vote.

The findings also highlight that larger models do not necessarily produce better results. For instance, the smallest model \texttt{phi} outperformed larger models in terms of recall. This suggests that model size alone is not a reliable indicator of performance for specific tasks, such as identifying cryptographically relevant software packages. Even models from online providers did not consistently outperform local models, indicating that with proper optimization, local LLMs can be competitive.

Another critical factor is the independence of models. As discussed in \Cref{sec:results:llm_independence}, correlated responses among models can diminish the effectiveness of the majority vote. Therefore, selecting heterogeneous models with diverse architectures and training data can help mitigate this issue.

The iterative approach of this study, which is based on validating a well selected labeled dataset, was essential for refining the preliminary results. We were able to significantly enhance the classification accuracy by continuously evaluating and optimizing prompts, parsing logic, and model selection based on performance metrics.

However, this approach has its limitations. Reliance on package descriptions and dependencies may result in the misclassification of packages with sparse or ambiguous publicly available information. Additionally, while the manual assessment used for verification is comprehensive, it may not capture all cryptographically relevant details.

\section{Conclusion and Outlook}
\label{sec:conclusion}


In summary, this study demonstrates that local LLMs can effectively identify cryptographically relevant software packages when properly optimized and combined. Our approach assists organizations in analyzing their software stacks for cryptographic relevance while preserving privacy within an on-premises, offline LLM framework. Our findings underscore the importance of prompt engineering, model independence, and an iterative optimization approach. Future work will focus on further enhancing prompt strategies and extending the detection method to further extract specific cryptographic primitives and generate comprehensive CBOMs. A prototype integrating our approach into a broader cryptographic asset discovery framework is under development and will be released via GitHub---see \cite{OthamiquasyDetectingCryptoPackages2026}.



\bibliographystyle{apalike}
{\small
\bibliography{misc/bibliography}}




\end{document}

%% file: misc/misc.tex
\makeatletter
\AtBeginDocument{%
\renewcommand{\ref}[1]{%
  \@latex@error{\string\ref\space nicht verwenden. Bitte durch cleveref (z.B. \string\Cref) ersetzen!}
}}
\makeatother

\newcommand{\ollama}[0] {\mbox{\textit{Ollama}}\xspace}
\newcommand{\gptFall}[0]{\mbox{\textit{GPT4All}}\xspace}

\newcommand{\eg}{e.g.,\xspace}

\usepackage{tikz}

\newcommand{\numref}[1]{%
  \tikz[baseline=(n.base)]\node (n) [draw,circle,inner sep=0.5pt,font=\small] {#1};%
}

%% file: images/approach.tex
\begin{tikzpicture}[
    node distance=7mm and 1cm,
    auto,
    start chain=going right,
    task/.style={
        rectangle,
        rounded corners,
        draw=black,
        fill=white,
        align=center,
        minimum width=2cm,
        minimum height=1cm,
    },
    highlight/.style={
        task,
        minimum width=2cm,
        draw=darkgray,
        fill=darkgray,
        text=white,
        font=\bfseries
    },
    arrow/.style={
        thick,
        ->,
        >=Stealth
    },
    number/.style={
        draw,
        circle,
        fill=white,
        inner sep=1pt,
        font=\small
    }
]

    \node[task] (package_list) {Packages\\List};
    \node[number, left=3pt of package_list.north west, anchor=center] {1};

    \node[task, right=of package_list] (prompt_engineering) {Prompt\\Engineering};
    \node[number, left=3pt of prompt_engineering.north west, anchor=center] {2};

    \node[task, right=of prompt_engineering] (query_llms) {Query\\LLMs};
    \node[number, left=3pt of query_llms.north west, anchor=center] {3};

    \node[highlight, below=of prompt_engineering] (validation) {Selection \&\\Validation};
    \node[number, left=3pt of validation.north east, anchor=center] {6};
    \node[number, left=3pt of validation.south east, anchor=center] {8};

    \node[task, left=of validation] (labels) {Labeled\\Data};
    \node[number, left=3pt of labels.north west, anchor=center] {7};

    \node[highlight, below=of validation] (majority_vote) {Majority\\Vote};
    \node[number, left=3pt of majority_vote.north west, anchor=center] {5};

    \node[task, right=of majority_vote] (aggregation) {Aggregation};
    \node[number, left=3pt of aggregation.north west, anchor=center] {4};

    \node[task, left=of majority_vote] (results) {Results\\Export};
    \node[number, left=3pt of results.north west, anchor=center] {9};

    \node[right=0.3cm of validation] {
        \protect\fontsize{32}{0}\selectfont\faSync
    };

    \draw[arrow] (package_list) -- (prompt_engineering);
    \draw[arrow] (prompt_engineering) -- (query_llms);
    \draw[arrow] (query_llms) -- (aggregation);
    \draw[arrow] (aggregation) -- (majority_vote);
    \draw[arrow] (majority_vote) -- (validation);
    \draw[arrow] (validation) -- (prompt_engineering);
    \draw[arrow] (validation) -- (results);
    \draw[arrow] (labels) -- (validation);

    \draw[arrow] (validation.north west) to [bend left=-45] (labels.north east);

\end{tikzpicture}

%% file: images/chart-empirical-vs-binom-vs-betabinom.tex
\begin{tikzpicture}
  \begin{axis}[
      ybar,
      width=15cm,
      height=10cm,
      ymajorgrids,
      grid=both,
      xlabel={Count True-Votes},
      ylabel={Packets},
      xtick={0,1,2,3,4,5},
      legend pos=north east,
      legend style={cells={anchor=west}},
      tick scale binop={\times},
      bar width=14pt,
      enlarge x limits=0.08,
      enlarge y limits={upper,value=0.17},
      ymin=0,
      nodes near coords,
      point meta=y,
      every node near coord/.append style={
          font=\small,
          rotate=90,
          anchor=west,
          color=black
      },
    ]

    \addplot[ybar, fill=black, draw=black] coordinates {
        (0,51965) (1,5597) (2,1885) (3,1215) (4,1198) (5,1669)
      };
    \addlegendentry{Empirical frequency};

    \addplot[ybar, fill=lightgray, draw=black] coordinates {
        (0,52697) (1,4091) (2,2293) (3,1677) (4,1404) (5,1367)
      };
    \addlegendentry{Beta-Bin($n=5$, $\hat \alpha=0$.$0751, \hat \beta=0$.$8368$)};

    \addplot[ybar, fill=white, draw=black] coordinates {
        (0,41345) (1,18545) (2,3327) (3,298) (4,13) (5,0)
      };
    \addlegendentry{Bin($n=5$, $\hat p=0.082321$)};

  \end{axis}
\end{tikzpicture}